\documentclass[prd,superscriptaddress,amsfonts,amssymb,amsmath,showpacs,twocolumn]{revtex4-2}

\usepackage{bm}
\usepackage{amsfonts}
\usepackage{latexsym}
\usepackage[latin1]{inputenc}
\usepackage{graphicx}
\usepackage{amsmath}
\usepackage{palatino}
\usepackage{mathpazo}
\usepackage{textcomp}
\usepackage{float}
\usepackage{booktabs}
\usepackage{dcolumn}
\usepackage{ragged2e}
\usepackage{hyperref}
\hypersetup{colorlinks,citecolor=blue}
\hypersetup{colorlinks=true,linkcolor=red,filecolor=magenta,    urlcolor=blue}
\usepackage{amsmath}
\usepackage{graphicx}% Include figure files

\usepackage{hyperref}% add hypertext capabilities
\usepackage[mathlines]{lineno}% Enable numbering of text and display math
\usepackage{amssymb}
\usepackage{enumitem}
\usepackage{mathtools}
\usepackage{mathrsfs}
\usepackage{setspace}
\usepackage{color}
\usepackage{graphicx}
\usepackage[dvipsnames]{xcolor}
\usepackage{mathalfa}
\usepackage{calligra}
\usepackage[misc]{ifsym}
\DeclareMathAlphabet{\mathpzc}{OT1}{pzc}{m}{it}
\DeclareMathAlphabet{\mathcalligra}{T1}{calligra}{m}{n}
\hypersetup{colorlinks=true,citecolor=blue,linkcolor=red,urlcolor=magenta}
\usepackage{xcolor}
\usepackage{orcidlink}
\usepackage[caption=false]{subfig}
\usepackage{commath}
\def\jnl@style{}
\def\aaref@jnl#1{{\jnl@style#1}}

\def\aaref@jnl#1{{\jnl@style#1}}

\def\aj{\aaref@jnl{AJ}}                   % Astronomical Journal
\def\apj{\aaref@jnl{ApJ}}                 % Astrophysical Journal
\def\apjl{\aaref@jnl{ApJ}}                % Astrophysical Journal, Letters
\def\apjs{\aaref@jnl{ApJS}}               % Astrophysical Journal, Supplement
\def\apss{\aaref@jnl{Ap\&SS}}             % Astrophysics and Space Science
\def\aap{\aaref@jnl{A\&A}}                % Astronomy and Astrophysics
\def\aapr{\aaref@jnl{A\&A~Rev.}}          % Astronomy and Astrophysics Reviews
\def\aaps{\aaref@jnl{A\&AS}}              % Astronomy and Astrophysics, Supplement
\def\mnras{\aaref@jnl{Mon.~Not.~Roy.~Astron.~Soc.}}             % Monthly Notices of the RAS
\def\prd{\aaref@jnl{Phys.~Rev.~D}}        % Physical Review D
\def\plb{\aaref@jnl{Phys.~Lett.~B}}        % Physics Letters B
\def\prc{\aaref@jnl{Phys.~Rev.~C}}  % Physical Review C
\def\prl{\aaref@jnl{Phys.~Rev.~Lett.}}    % Physical Review Letters
\def\qjras{\aaref@jnl{QJRAS}}             % Quarterly Journal of the RAS
\def\skytel{\aaref@jnl{S\&T}}             % Sky and Telescope
\def\ssr{\aaref@jnl{Space~Sci.~Rev.}}     % Space Science Reviews
\def\zap{\aaref@jnl{ZAp}}                 % Zeitschrift fuer Astrophysik
\def\nat{\aaref@jnl{Nature}}              % Nature
\def\aplett{\aaref@jnl{Astrophys.~Lett.}} % Astrophysics Letters
\def\apspr{\aaref@jnl{Astrophys.~Space~Phys.~Res.}} % Astrophysics Space Physics Research
\def\physrep{\aaref@jnl{Phys.~Rep.}}      % Physics Reports
\def\physscr{\aaref@jnl{Phys.~Scr}}       % Physica Scripta
\def\commat{\aaref@jnl{Comm.~Math.~Phys.}}              % Communications in Mathematical Physics
\def\science{\aaref@jnl{Science}}               % Science
\def\cqg{\aaref@jnl{Classical Quant.~Grav.}}            % Classical and Quantum Gravity
\def\jpcs{\aaref@jnl{JPCS}}                                     % Journal of Physics Conference Series
\def\ijmpd{\aaref@jnl{Int.~J.~Mod.~Phys.~D}}                    % International Journal of Modern Physics D
\def\grg{\aaref@jnl{Gen.~Relat.~Gravit.}}               % General Relativity and Gravitation
\def\rpp{\aaref@jnl{Rep.~Prog.~Phys.}}          % Reports on Progress in Physics
\def\npa{\aaref@jnl{Nucl.~Phys.~A}}        % Nuclear Physics A
\def\lrr{\aaref@jnl{Living Rev.~Rel.}}                   % Living reviews in relativity
\def\jcap{\aaref@jnl{J.~Cosmology Astropart.~Phys.}}    % Journal of cosmology and astroparticle physics
\def\rmp{\aaref@jnl{Rev.~Mod.~Phys.}}   %Reviews of modern physics
\def\epjc{\aaref@jnl{Eur.~Phys.~J.~C}}

\allowdisplaybreaks[1]
\begin{document}

\preprint{APS/123-QED}

\title{Constraining Anisotropic Cosmological Model in $\mathpzc{f}(\mathcal{R},\mathscr{L}_m)$ Gravity}% Force line breaks with \\

\author{N. S. Kavya\orcidlink{0000-0001-8561-130X}}
\email{kavya.samak.10@gmail.com}
\affiliation{Department of P.G. Studies and Research in Mathematics,
 \\
 Kuvempu University, Shankaraghatta, Shivamogga 577451, Karnataka, INDIA
}%
 
\author{V. Venkatesha\orcidlink{0000-0002-2799-2535}}%
 \email{vensmath@gmail.com}
\affiliation{Department of P.G. Studies and Research in Mathematics,
 \\
 Kuvempu University, Shankaraghatta, Shivamogga 577451, Karnataka, INDIA
}%


\author{Sanjay Mandal\orcidlink{0000-0003-2570-2335}}%
\email{sanjaymandal960@gmail.com}
\affiliation{
 Department of Mathematics, Birla Institute of Technology and Science-Pilani,\\
 Hyderabad Campus, Hyderabad 500078, INDIA
}%

\author{P.K. Sahoo\orcidlink{0000-0003-2130-8832}}
\email{pksahoo@hyderabad.bits-pilani.ac.in}
\affiliation{
 Department of Mathematics, Birla Institute of Technology and Science-Pilani,\\
 Hyderabad Campus, Hyderabad 500078, INDIA
}%


\date{\today}% It is always \today, today,

\begin{abstract}
The observational evidence regarding the present cosmological aspects tells us about the presence of very little anisotropy in the universe on a large scale. Here, in this paper, we attempt to study locally rotationally symmetric (LRS) homogeneous Bianchi-I spacetime with the isotropic matter distribution. This is done within the framework of $\mathpzc{f}(\mathcal{R},\mathscr{L}_m)$ gravity. Particularly, we consider a non-linear $\mathpzc{f}(\mathcal{R},\mathscr{L}_m)$ model, $\mathpzc{f}(\mathcal{R},\mathscr{L}_m)=\dfrac{1}{2}\mathcal{R}+\mathscr{L}_m^{\,{\alpha}}$. Furthermore, $\omega$, the equation of state parameter, which is vital stuff in determining the present phase of the universe is constrained. To constrain the model parameters and the equation of state parameter, we use 57 Hubble data points and 1048 Pantheon supernovae type Ia data sample. And, for our statistical analysis, we use Markoc Chain Monte Carlo (MCMC) simulation. Moreover, with the help of obtained values of parameters, we measure the anisotropy parameter for our model.    
\begin{description}
\item[Keywords]
Equation of state parameter, $\mathpzc{f}(\mathcal{R},\mathscr{L}_m)$ gravity, observational constraints, anisotropy\\ parameter.

\end{description}
\end{abstract}

\maketitle

%\tableofcontents
\section{INTRODUCTION}\label{sectionI}
		\par Over the past few decades, many scientific explorations have been taking place to decipher the mystic behavior of the universe. Right from the early time inflation to the late time acceleration, from the black holes to the wormholes, from the dark energy to the gravitational waves, their entire course has been probing the very nature of the universe. Just to look into the cosmological principle, the universe on a large scale, was presumed to be both isotropic and homogeneous. But in 1992, Cosmic Background Explorer (COBE) successfully made a significant assertion about the existence of a small anisotropy in the large-scale cosmic microwave background \cite{1cobe}. Moreover, in the later years, this was further supported by the measurements made by Balloon Observations of Millimetric Extragalactic Radiation and Geophysics (BOOMERanG) \cite{1boomerang}, Cosmic Background Imager (CBI) \cite{cbi}, Wilkinson Microwave Anisotropy Probe (WMAP)\cite{1wmap}, and the Plank collaborations\cite{1planck}. Furthermore, intriguing advancements in the field of cosmology took place through the observational results of the two teams led by Perlmutter and Riess \cite{perlmutter,1riess}. These studies strive to endorse that the universe is currently in the phase of accelerated expansion. To this point, there arose a question regarding the isotropic nature of the expansion of the universe. Interestingly, recent developments suggest that the universe tends to expand at a different rate in different directions \cite{1planck2}. Though FLRW cosmology is most successful, it is built based on cosmological principles. However, the observational evidence attempts to elucidate the presence of a slight difference in the strengths of microwaves coming from different axes. For this reason, the spacetime that can appropriately describe anisotropic and homogeneous geometry is Bianchi cosmology. Several works on such Bianchi cosmology with different modified gravity frameworks can be found in the literature. (See ref \cite{b1,b2,b3,b4,b5,b6,b7,b8,b9,b10,b11, 2fr1})
	
		\par In the present scenario, to deal with the study of such aspects, the modified theoretic approach sounds more potent. Among these, the $\mathpzc{f}(\mathcal{R})$ theory of gravity has produced a reliable framework for evaluating the current cosmic evolution \cite{1fofr}. Indeed, $\mathpzc{f}(\mathcal{R})$ theories can adequately explain the interpretations of late-time acceleration \cite{1fr1,2fr1}, the exclusion of the dark matter entity in the analysis of the dynamics of massive test particles \cite{1fr2}, and the unification of inflation with dark energy \cite{1fr3}. Furthermore, numerous justifications indicate that the higher-order theories, like $\mathpzc{f}(\mathcal{R})$ gravity, are capable of explaining the flatness of galaxies' rotational curves \cite{1fr4}. With these motivations, several coupling theories came into existence \cite{1frt, 1frboxrt, 1frtrmunutmunu}. One such theory is the  $\mathpzc{f}(\mathcal{R},\mathscr{L}_m)$ theory of gravity \cite{1frlm}. Notably, this favors the occurrence of an extra force that is orthogonal to four-velocity. In addition, the so-called `extra' force accounts for the non-geodesic motions of the test particle. Consequently, a violation of the equivalence principle can be observed. Numerous contributions to this theory can be seen in the literature \cite{1solarsystem,1frlm1,1frlm2,1frlm3,1frlm4,1frlm5,ga,1frlm7,1frlm8}. Recently, Jaybhaye et al have studied cosmology in $\mathpzc{f}(\mathcal{R},\mathscr{L}_m)$ gravity \cite{1frlm6}.
		
		\par In the present work, we center on the study of theoretical exploration and observational validation of the LRS Bianchi type I spacetime and effectuate this in terms of $\mathpzc{f}(\mathcal{R},\mathscr{L}_m)$  formalism. Moreover, in assessing the expanding universe the equation of state parameter plays a prominent role. This predicts the fluid type in spacetime. In our work, we emphasize constraining this cosmological parameter $\omega$ and obtaining the best fit values as per the observational measurements. This is accomplished with a statistical approach for incorporated sets of data samples. We use two types of data samples such as Hubble measurements and Pantheon SNe Ia sample . Further, with the anisotropy parameter, we measure anisotropy in spacetime. 
		
		\par This manuscript is organized as follows: in section \ref{sectionII}, the basic formulation of $\mathpzc{f}(\mathcal{R},\mathscr{L}_m)$ gravity is presented. The analysis of LRS Bianchi I within the framework of the $\mathpzc{f}(\mathcal{R},\mathscr{L}_m)$ gravity is made in section \ref{sectionIII}. Section \ref{sectionIV} is brought with the examination of observational constraints and discussion of results. Finally, the last section \ref{sectionV}, gives some concluding remarks.

%%%%%%%%%%%%%%%%%%%%%%%%%%%%%%%%%%%%%%%%%%%%%%%%%%%%%%%%%%%%%%%%%%%%%%%%%%%
	
		\section{THE BASIC FIELD EQUATIONS IN $\mathpzc{f}(\mathcal{R},\mathscr{L}_m)$ GRAVITY}\label{sectionII}			
			
		\par With the matter lagrangian density $\mathscr{L}_m$ and the Ricci scalar $\mathcal{R}$, the action integral for $ \mathpzc{f}(\mathcal{R},\mathscr{L}_m)$ theory reads,
		\begin{equation}\label{action}
			S=\int \mathpzc{f}(\mathcal{R},\mathscr{L}_m)	\sqrt{-g}\, d^4x,
		\end{equation}
		where $\mathpzc{f}$ represents an arbitrary function of $\mathcal{R}$ and $\mathscr{L}_m$. 
		
		\par The field equation for the $\mathpzc{f}(\mathcal{R},\mathscr{L}_m)$ gravity \cite{1frlm}, obtained by varying the action integral \eqref{action} with respect to the metric tensor $g_{\mu\nu}$ is given by, 
	\textbf{	\begin{equation}\label{fieldequation1}
			\begin{split}
				\mathpzc{f}_\mathcal{R}(\mathcal{R},\mathscr{L}_m)\mathcal{R}_{\mu\nu}+(g_{\mu\nu}\nabla_\mu\nabla^{\mu}-\nabla_\mu\nabla_\nu)\mathpzc{f}_\mathcal{R}(\mathcal{R},\mathscr{L}_m)\\-\dfrac{1}{2}\left[\mathpzc{f}(\mathcal{R},\mathscr{L}_m)- \mathpzc{f}_{\mathscr{L}_m}(\mathcal{R},\mathscr{L}_m)\mathscr{L}_m\right]g_{\mu\nu}=\\\dfrac{1}{2}\mathpzc{f}_{\mathscr{L}_m}(\mathcal{R},\mathscr{L}_m)\mathcal{T}_{\mu\nu}.
			\end{split}
		\end{equation}}
		Here, $\mathpzc{f}_\mathcal{R}(\mathcal{R},\mathscr{L}_m)\equiv\frac{\partial \mathpzc{f}(\mathcal{R},\mathscr{L}_m)}{\partial \mathcal{R}}$, $\mathpzc{f}_{\mathscr{L}_m}(\mathcal{R},\mathscr{L}_m)\equiv\frac{\partial \mathpzc{f}(\mathcal{R},\mathscr{L}_m)}{\partial \mathscr{L}_m}$, and $\mathcal{T}_{\mu\nu}$ is the Energy-Momentum Tensor (EMT) that can be expressed as,
		\begin{equation}
			\mathcal{T}_{\mu\nu}=-\dfrac{2}{\sqrt{-g}} \dfrac{\delta(\sqrt{-g}\mathscr{L}_m)}{\delta g^{\mu\nu}}=g_{\mu\nu}\mathscr{L}_m-2\dfrac{\partial \mathscr{L}_m}{\partial g^{\mu\nu}}.
		\end{equation}  
		\par Now, from the explicit form of the field equation \eqref{fieldequation1}, the covariant divergence of EMT $\mathcal{T}_{\mu\nu}$ can be obtained as, 
		\begin{equation}
			\nabla^\mu \mathcal{T}_{\mu\nu}=2\left\lbrace \nabla^\mu \text{ln}\left[\mathpzc{f}_{\mathscr{L}_m}(\mathcal{R},\mathscr{L}_m) \right]\right\rbrace \dfrac{\partial \mathscr{L}_m }{\partial g^{\mu\nu}}. 
		\end{equation}
		\par Furthermore, on contracting the field equation \eqref{fieldequation1} we get,
		\begin{equation}\label{traceoffieldequation}
			\begin{split}
				3\nabla_\mu\nabla^{\mu}\mathpzc{f}_\mathcal{R}(\mathcal{R},\mathscr{L}_m)+\mathpzc{f}_\mathcal{R}(\mathcal{R},\mathscr{L}_m)\mathcal{R}-2\left[\mathpzc{f}(\mathcal{R},\mathscr{L}_m)\right.\\\left. -\mathpzc{f}_{\mathscr{L}_m}(\mathcal{R},\mathscr{L}_m)\mathscr{L}_m\right]=\dfrac{1}{2}\mathpzc{f}_{\mathscr{L}_m}(\mathcal{R},\mathscr{L}_m)\mathcal{T}.
			\end{split}
		\end{equation}
		\par By considering the above equation, the relation between the trace of EMT $\mathcal{T}=\mathcal{T}^\mu_\mu$, $\mathscr{L}_m$ and  $\mathcal{R}$ can be established.
		
%%%%%%%%%%%%%%%%%%%%%%%%%%%%%%%%%%%%%%%%%%%%%%%%%%%%%%%%%%%%%%%%%%%%%%%%%%%%
		\section{LRS BIANCHI-I COSMOLOGY IN $\mathpzc{f}(\mathcal{R},\mathscr{L}_m)$ GRAVITY}\label{sectionIII}
		\par For anisotropic and spacially homogeneous LRS Bianchi-I spacetime, the metric is described by, 
		\begin{equation}
			ds^2=-dt^2+\mathcal{A}^2(t)\;dx^2+\mathcal{B}^2(t)\;dy^2+\mathcal{B}^2(t)\;dz^2,
		\end{equation}
		where, $\mathcal{A}$ and $\mathcal{B}$ are metric potentials that are the functions of (cosmic) time $t$ alone. If $\mathcal{A}(t)=\mathcal{B}(t)=a(t)$, then one can analyze the scenarios in flat FLRW spacetime.
		Now, the Ricci scalar for LRS Bianchi-I spacetime can be expressed as,
		\begin{equation}
			R=2\left[\dfrac{\ddot{\mathcal{A}}}{\mathcal{A}}+\dfrac{2\ddot{\mathcal{B}}}{\mathcal{B}}+\dfrac{2\dot{\mathcal{A}}\dot{\mathcal{B}}}{\mathcal{A}\mathcal{B}}+\dfrac{\dot{\mathcal{B}}^2}{\mathcal{B}^2}\right]
		\end{equation}
		
		With the directional Hubble parameters $H_x$, $H_y$ and $H_z$, the Ricci scalar for the corresponding metric is given by,
		
		\begin{equation}
			R=2(\dot{H}_x+2\dot{H}_y)+2(H_x^2+3H_y^2)+4H_xH_y.
		\end{equation}
		
		Here, $H_x=\frac{\dot{\mathcal{A}}}{\mathcal{A}}$ and $H_y= \frac{\dot{\mathcal{B}}}{\mathcal{B}}=H_z$ indicate the directional Hubble parameters along the corresponding coordinate axes. For $H_x=H_y=H$, i.e., for FLRW cosmology, the equation $R=2(\dot{H}_x+2\dot{H}_y)+2(H_x^2+3H_y^2)+4H_xH_y$ reduces to $R=6(2 H^2+\dot{H})$. In the present work, we are supposing the matter distribution to be described by the energy-momentum tensor of a perfect fluid,
		\begin{equation}
		    \mathcal{T}_{\mu\nu}=(\rho+p)U_{\mu}U_{\nu}+p\, g_{\mu\nu},
		\end{equation}
		where $\rho$ is the energy density and $p$ is the pressure. The four-velocity, $U_{\mu}$ satisfies the condition $U_{\mu}U^{\mu}=-1$ and $U_{\mu}U^{\mu};_{\nu}=0$. Thus, the field equation \eqref{fieldequation1} takes the form,
		
		\begin{align}
		\label{fe1}\begin{split}
			-\dot{\mathpzc{f}}_\mathcal{R}(H_{x}+2 H_{y})+(\dot{H}_{x}(t)+2 \dot{H}_{y}+H_{x}^{2}+2 H_{y}^{2})\, \mathpzc{f}_\mathcal{R}\\-\dfrac{1}{2}\left(\mathpzc{f}-\mathpzc{f}_{\mathscr{L}_m} \mathscr{L}_m\right)=-\dfrac{\rho \mathpzc{f}_{\mathscr{L}_m}}{2},&
		\end{split}\\
		\label{fe2}\begin{split}
			-\ddot{\mathpzc{f}}_\mathcal{R}-2 H_{y}\, \dot{\mathpzc{f}}_\mathcal{R}+(\dot{H}_{x}+H_{x}^{2}+2 H_{x} H_{y}) \mathpzc{f}_\mathcal{R}\\-\dfrac{1}{2}\left(\mathpzc{f}-\mathpzc{f}_{\mathscr{L}_m} \mathscr{L}_m\right)=\dfrac{p \mathpzc{f}_{\mathscr{L}_m}}{2},&
		\end{split}\\
		\label{fe3}\begin{split}
			-\ddot{\mathpzc{f}}_\mathcal{R}-2 H_{x}\, \dot{\mathpzc{f}}_\mathcal{R}+(\dot{H}_{y}+2 H_{x} H_{y}) \mathpzc{f}_\mathcal{R}\\-\dfrac{1}{2}\left(\mathpzc{f}-\mathpzc{f}_{\mathscr{L}_m} \mathscr{L}_m\right)=\dfrac{p \mathpzc{f}_{\mathscr{L}_m}}{2}.&
		\end{split}
		\end{align}
		The dot $(\cdot)$ here represents the derivative with respect to the time $t$ and $\mathpzc{f}\equiv \mathpzc{f}(\mathcal{R},\mathscr{L}_m)$.
		
		\par Further, one can express the spatial volume $\mathcal{V}$ of the spacetime as,  
		\begin{equation}
			\mathcal{V}=a^3=AB^2.
		\end{equation}
		Thus the mean value of the Hubble parameter is given by,
		\begin{align}
			H=\dfrac{\dot{a}}{a}=\dfrac{1}{3}(H_x+2H_y).
		\end{align} 
	In further study, we are going to investigate the physical cosmological model and their application in the context of $\mathpzc{f}(\mathcal{R},\mathscr{L}_m)$ gravity using the above set of equations.	
		\subsection{Physical Model:}
		\par In the present study, we shall focus on the cosmological aspects of $\mathpzc{f}(\mathcal{R},\mathscr{L}_m)$ theory, with the relation between $\mathcal{R}$ and $\mathscr{L}_m$ being
 		\begin{equation}
			\mathpzc{f}(\mathcal{R},\mathscr{L}_m)=\dfrac{1}{2}\mathcal{R}+\mathscr{L}_m^{\,{\alpha}}	,
		\end{equation}
		where, $\alpha\ne 0$ is a model parameter and one can retain GR for $\alpha=1$.
		
		\par Now, to find an exact solution to the field equations \eqref{fe1}-\eqref{fe3}, we have to consider the constraining relation. To this point, we shall presume the anisotropic relation that can be written in terms of shear $(\sigma)$ and expansion scalar $(\theta)$ as, 
		$$\theta^2 \propto \sigma^2, $$
		so that, for constant $\frac{\sigma}{\theta}$, the Hubble expansion can achieve isotropy \cite{m2,m3}. This condition gives rise to
		\begin{equation}
		    \mathcal{A}(t)=\mathcal{B}(t)^n,
		\end{equation}
		for some real non-zero $n$, and for $n=1$, we can retrieve flat FLRW cosmology. With this, one can get the relation between directional Hubble parameters as, 
		\begin{equation}\label{hxnhy}
		    H_x=n H_y.
		\end{equation}
		Therefore, averaged Hubble parameter can takes the form,
		\begin{equation}\label{hhy}
		    H=\dfrac{n+2}{3} H_y.
		\end{equation}
		Additionally, we shall relate the the pressure $p$ and energy density $\rho$ by,
		\begin{equation}
			p=\omega\rho.
		\end{equation}
		Now, we have two choices for the Lagrangian to proceed further such as $\mathscr{L}_m=-\rho$ or $\mathscr{L}_m=p$ \cite{rho1}. But, in our study we consider $\mathscr{L}_m=-\rho$ because it is the most adequate choice presented in \cite{rho1}. In literature, these are many studies which have been explored the choices for $\mathscr{L}_m$ and their applications [to see more details please check the references therein \cite{rho1,rho2}.

		Applying the above conditions, the field equations \eqref{fe1}-\eqref{fe3} become,
		\begin{align}
			\dfrac{9(2n+1) }{(n+2)^{2}}H^{2}&=-(-\rho)^\alpha,\\
			\dfrac{6\dot{H}}{(n+2)}+\dfrac{27 H^2}{(n+2)^2}&=-(-\rho)^\alpha\left[\alpha(1+\omega)+1\right],\\
			3\dot{H}\left(\dfrac{n+1}{n+2}\right)+9H^2\dfrac{n^2+3}{(n+2)^2}&= - (-\rho)^\alpha \left[\alpha(1+\omega)+1\right].
		\end{align}
		
		\par With the help of aforementioned equations we can obtain an expression for the Hubble parameter $H$ in terms of redshift $z$ as,
		%\begin{equation}
		%    H(z)=\frac{4 \alpha n+8 \alpha-2 n-4}{(\frac{1}{z+1})^{\frac{6 \omega \alpha n-6 \alpha n+3 \omega \alpha+15 \alpha+6 n-6}{4 \alpha n+8 \alpha-2 n-4}}}
		%\end{equation}
		\begin{equation}
		    H(z)=\gamma_1\left(z+1\right)^{\gamma_1/\gamma_2},
		\end{equation}
		where, $\gamma_1=2(n+2)$ and $\gamma_2= -3[\alpha(\omega+1)(2n+1)+2(n-1)]$. Here, we used the scale factor $a(t)$ and redshift relation, which is given by
		\begin{equation}
		a(t)=\frac{1}{1+z}.
		\end{equation}
		
		We  now aim to constrain the model parameters ($\alpha,\,\, n$) and cosmological parameter ($\omega$) using various observational measurements. Doing this helps us to present a physically realistic cosmological model, which can obey the astrophysical observations.
%%%%%%%%%%%%%%%%%%%%%%%%%%%%%%%%%%%%%%%%%%%%%%%%%%%%%%%%%%%%%%%%%%%%%%%%%%%	
		\section{OBSERVATIONAL CONSTRAINTS AND VALIDATION OF THE RESULTS}\label{sectionIV}
		\par So far, we looked into the formulation of LRS Bianchi I cosmology in $\mathpzc{f}(\mathcal{R},\mathscr{L}_m)$ gravity. It is to be noted that the substantial validation of the values for the parameters is significant in analyzing the cosmological aspects. In this regard, the present section puts forth the observational interpretations of the current scenario. The statistical technique we adopted assists us to constrain the parameters such as, $\omega$, $\alpha$ and $n$. In particular, we have opted for the Markov Chain Monte Carlo (MCMC) with standard Bayesian technique. Further, with the pseudo-chi-squared function $\chi^2$, the probability function
		\begin{equation}
		    \mathcal{L}\propto e^{-\frac{\chi^2}{2}},
		\end{equation}
		provides the best fit values for the parameters.
		 \begin{widetext}   
		
		\begin{figure}[H]
			\centering
			\includegraphics[scale=0.88]{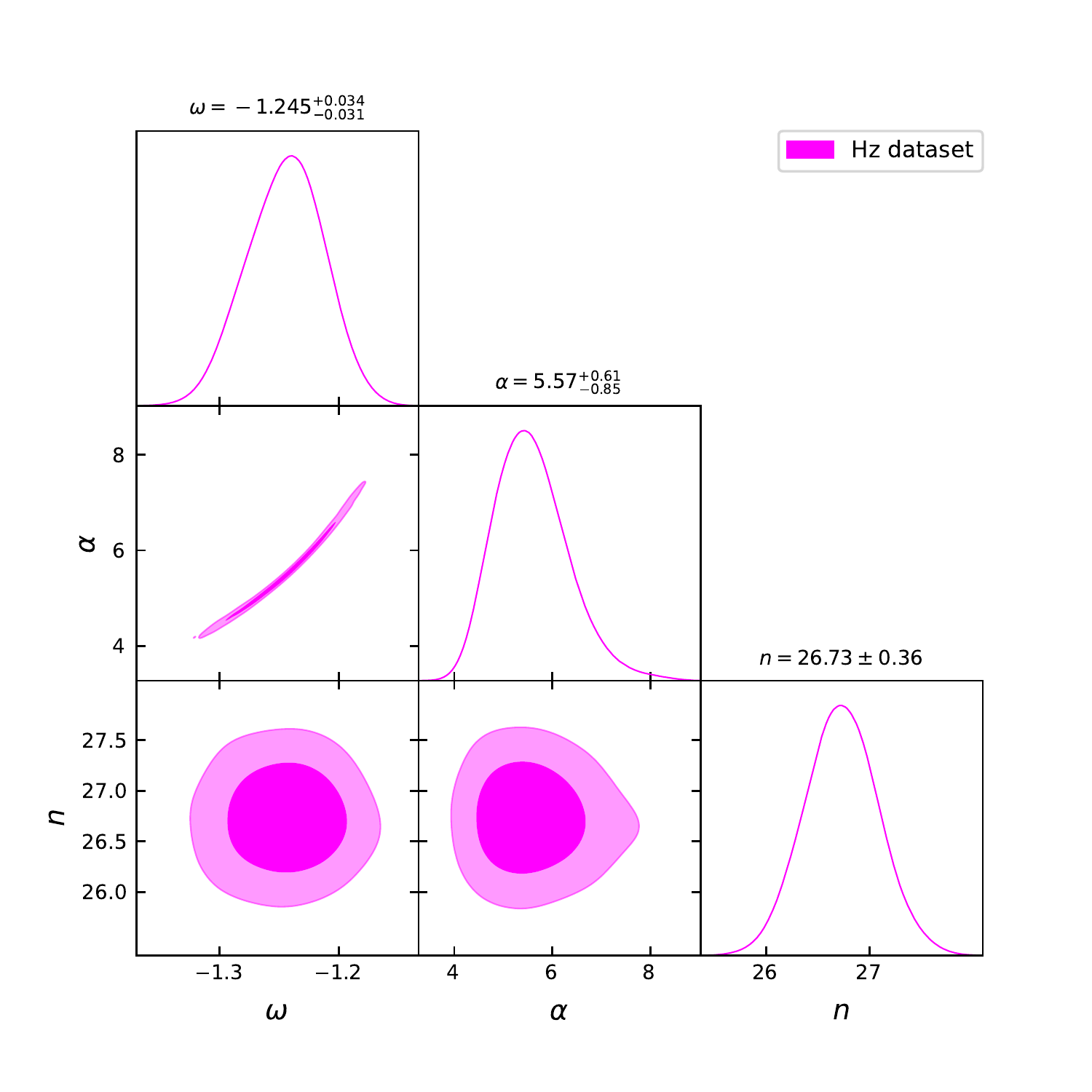}
			\caption{Contour plot with $1-\sigma$ and $2-\sigma$ errors for the parameters $\omega$, $\alpha$ and $n$ along with the constraint values for Hubble dataset.}
			\label{fig:kh}
		\end{figure}
		
		\end{widetext}
		 Presently, to achieve this, we center on two datasets namely, $H(z)$ and Pantheon data. To proceed further, we consider the priors on parameters, which are $(-2.0<\omega<1.0)$ to keep in mind all the possible cases of equation of state parameter, $(-10<\alpha<+ 10)$ as it is a free model parameter, and $(0<n<100)$ to measure the anisotropy and also to keep in mind that our model should fit the observational dataset as well.
		\subsection{H(z) dataset}
		    \par The prominence of the analysis of the Hubble parameter lies in the exploration of the expanding universe. Moreover, this can be expressed in terms of the redshift parameter z, which is quite useful in many circumstances. At some particular redshifts, we can infer the value of the Hubble parameter. To this end, the determination of its value from line-of-sight BAO data is one of the most successful techniques. Furthermore, another widely used approach to find H(z) is the differential age method. In the redshift range $0.07\le z\le  2.41$, the 31 $H(z)$ points obtained from differential age method \cite{hz1,hz2,hz3,hz4,hz5,hz6,hz7} and 26 points from other methods including BAO (see ref \cite{hz8,hz9,hz10,hz11,hz12,hz13,hz14,hz15,hz16,hz17,hz18,hz19}), provide 57 points of the $H(z)$ dataset. Now, as mentioned earlier, we consider the pseudo chi-square function $\chi^2$ to evaluate the unknown parameters. For $H(z)$ dataset, it is given by,
            \begin{equation}
             \chi^2_{Hz}(\omega,\alpha,n)=\sum\limits_{k=1}^{57}\dfrac{\left[H_{th}(z_k,\omega,\alpha,n)-H_{ob}(z_k)\right]^2}{\sigma^2_{H(z_k)}}.
            \end{equation}
            Here, $H_{th}$ indicates the theoretically obtained value of the Hubble parameter and $H_{ob}$ represents its observed value and $\sigma$ is the standard deviation.

		\begin{widetext}
		
		\begin{figure}[H]
			\centering
			\includegraphics[scale=0.6]{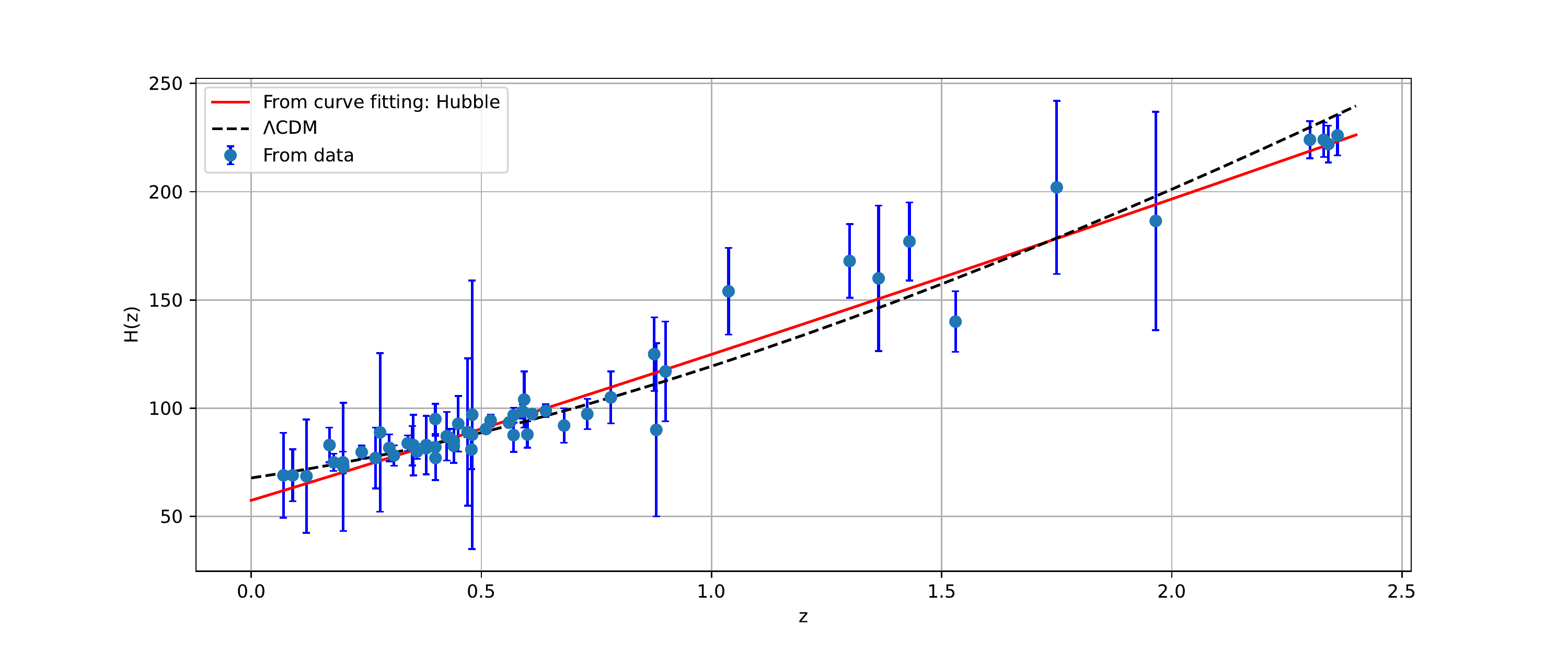}
			\caption{The profile of Hubble parameter versus redshift $z$. The line in red shows the curve for the model and the dotted line in black represent the $\Lambda$CDM model with $\Omega_{m0}$, $\Omega_{\Lambda 0}$ having the values 0.3 and 0.7 respectively. The dots with error bars in blue depicts 57 $H(z)$ sample points.
			}
			
			\label{fig:khf}
		\end{figure}
		
		\end{widetext}
		
		\subsection{Pantheon dataset}
		    \par The SNe Ia holds a central role in explaining the expanding universe. Significantly, the spectroscopically collected SNe Ia data such as, SuperNova Legacy Survey (SNLS), Sloan Digital Sky Survey (SDSS),Hubble Space Telescope (HST) survey, Panoramic Survey Telescope and Rapid Response System(Pan-STARRS1) provide a solid evidence in this regard. The recent data sample of the SNe Ia, the Pantheon dataset comprises 1048 magnitudes for the distance modulus  estimated over the range of $0.01\le z\le 2.3$ for the redshift z \cite{pantheon}. In order to find the best fits for the model in hand, we perform the analogical assessment between the theoretical and observational values of the distance moduli $\mu(z_k)$. Theoretically,
		    \begin{equation}
		        \mu^{th}(z_k)=\mu_0+5\,\text{log}_{10}(\mathcal{D}_L(z_k)),
		    \end{equation}
		    with the nuisance parameter
		    \begin{equation}
		        \mu_0= 25+5\, \text{log}_{10}\left(\dfrac{1}{H_0 Mpc}\right),
		    \end{equation}
		    and the luminosity distance
		    \begin{equation}
		        \mathcal{D}_L(z)=(1+z)\int\limits_0^z \dfrac{c}{H(\xi)}d\xi.
		    \end{equation}
		Here, we take $H_0=69$ km/s/Mpc \cite{patheon}. 
		Now, $\chi^2$ function for Pantheon data sample, with covariance metric $\mathcal{C}_{SNe}$ is given by,
		\begin{equation}
		    \chi^2_{SNe}(\mu_0,\omega,\alpha,n)=\sum\limits_{k,l=1}^{1048}\bar{\mu}_k \left(\mathcal{C}_{SNe}^{-1}\right)_{kl}\bar{\mu}_l,
		\end{equation}
		where, $\bar{\mu}_k=\mu^{th}(z_k,\omega,\alpha,n)-\mu^{ob}(z_k)$.
		\begin{widetext}
		
		\begin{figure}[H]
			\centering
			\includegraphics[scale=0.9]{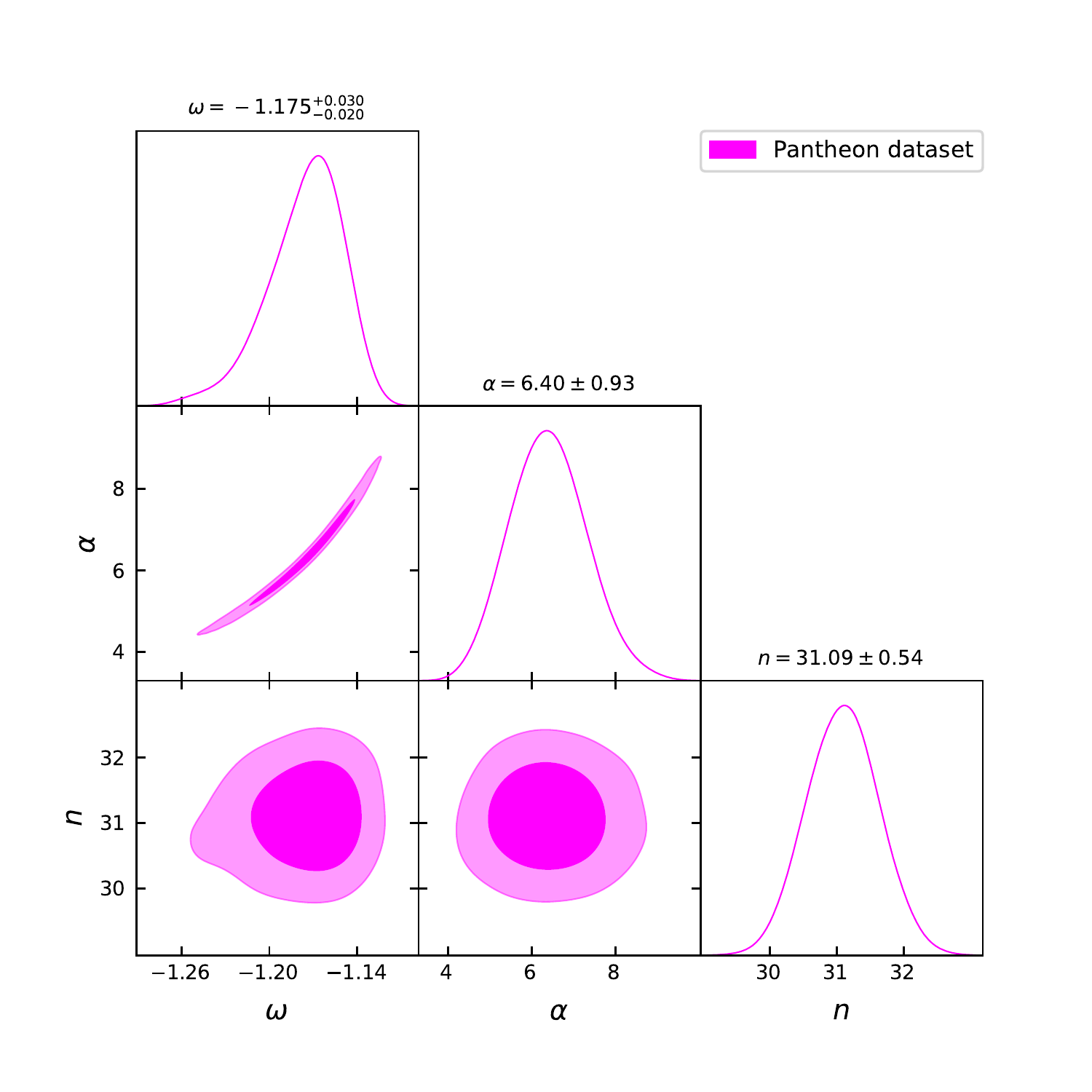}
			\caption{Contour plot with $1-\sigma$ and $2-\sigma$ errors for the parameters $\omega$, $\alpha$ and $n$ along with the constraint values for pantheon dataset.}
			\label{fig:kp}
		\end{figure}
		
		\begin{figure}[H]
			\centering
	   	\includegraphics[width=1.0\linewidth]{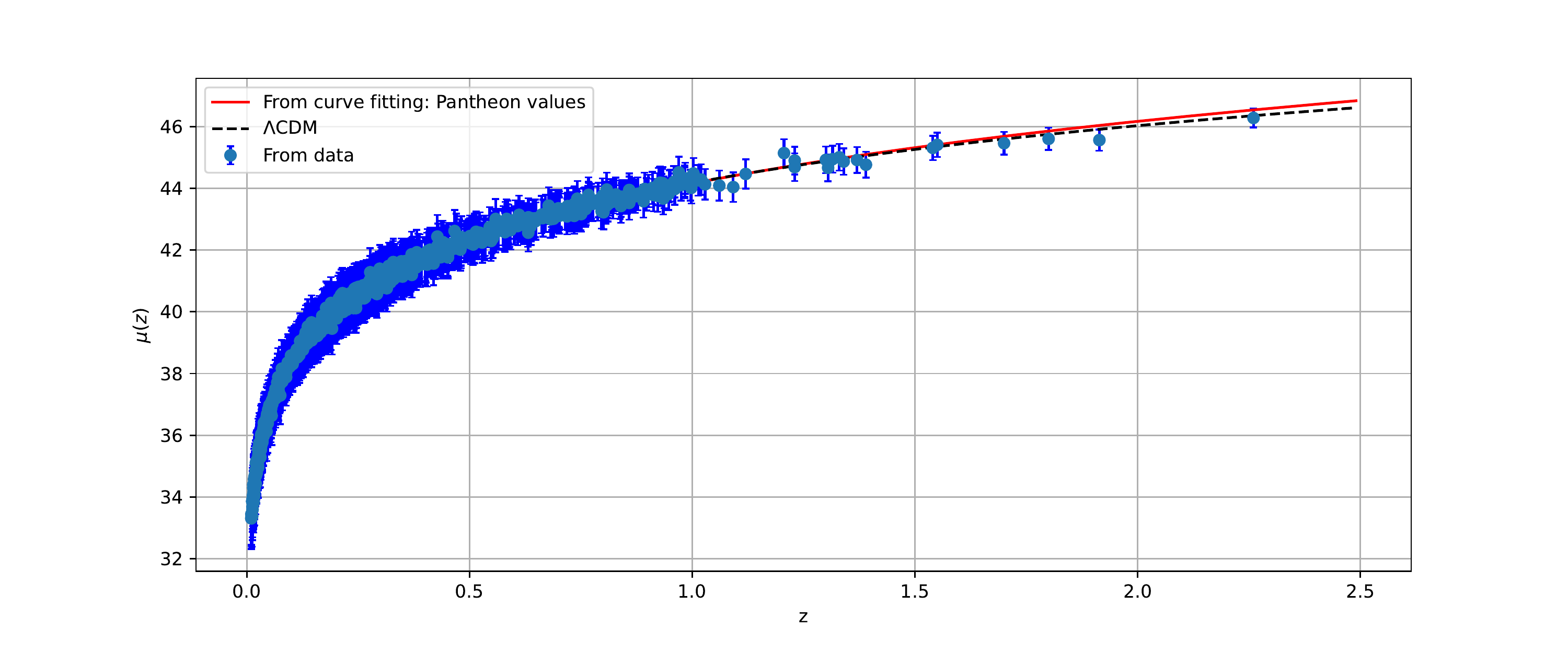}
			\caption{The profile of distance modulus versus redshift $z$. The line in red shows the curve for the model and the dotted line in black represent the $\Lambda$CDM model with $\Omega_{m0}$, $\Omega_{\Lambda 0}$ with the values 0.3 and 0.7, respectively. The dots with error bars in blue depicts 1048 pantheon sample points.}
			\label{fig:kpf}
		\end{figure}
		
		\end{widetext}
		
		\subsection{H(z)+Pantheon dataset}
		 \par In the previous sections, we dealt with two kinds datasets, $H(z)$ and SNe Ia. In this section, we shall consider the combination of these two sets of sample points. By the same token, to examine the best fits for our model with H(z)+Pantheon dataset, we consider $\chi^2$ function as given below.
		 \begin{equation}
		     \chi^2_{\text{comb}}=\chi^2_{Hz}+\chi^2_{SNe}.
		 \end{equation}
		 	\subsection{Anisotropy Parameter}
        \par It is known that, the universe is expanding at a different pace in different directions. This certainly leads to the anisotropy in the geometric structure of spacetime. A physical quantity that measures the amount of anisotropy that arises due to the expanding universe is the anisotropic parameter. Mathematically, this can be expressed as, 
        \begin{equation}
    		\Delta=\dfrac{1}{3}\sum\limits_{i=1}^3\left(\dfrac{H_i-H}{H} \right)^2.
		\end{equation}
        For the present problem, this takes the form: 
        \begin{equation}
		\Delta=\dfrac{2}{9H^2}\left(H_x-H_y\right)^2.
		\end{equation}
        From \eqref{hxnhy} and \eqref{hhy} we have, 
        \begin{equation}
		\Delta=2\left(\dfrac{n-1}{n+2}\right)^2.
		\end{equation}
        In the previous section, from the distinct observational datasets, we have obtained the value for n. Now, the corresponding value of the anisotropy measure $\Delta$ is given in the table \ref{tab:my_label}.
		 \begin{widetext}
		 
		\begin{figure}[H]
			\centering
			\includegraphics[scale=1.2]{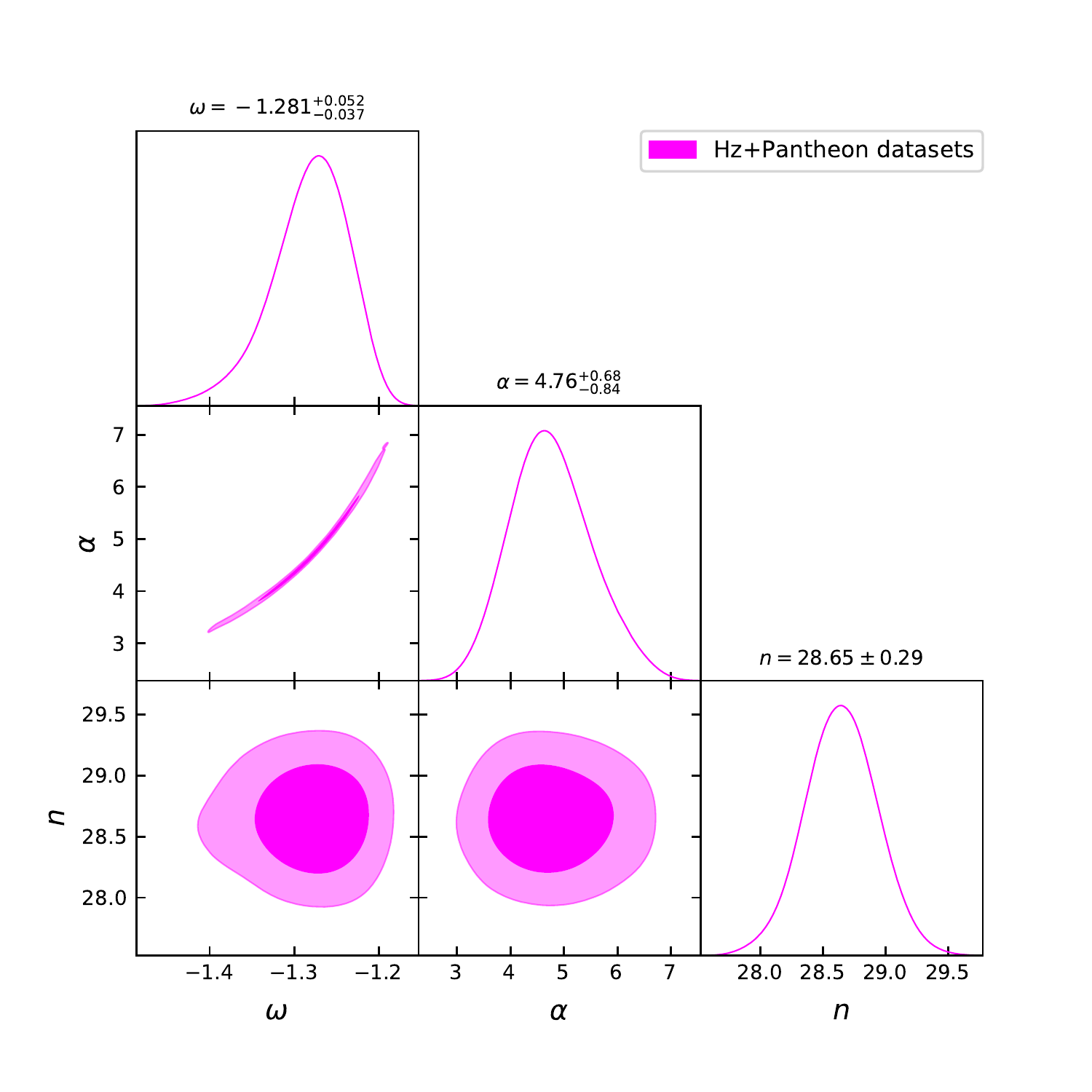}
			
			\caption{Contour plot with $1-\sigma$ and $2-\sigma$ errors for the parameters $\omega$, $\alpha$ and $n$ along with the constraint values for $H(z)+$Pantheon dataset.}
			\label{fig:Kcombine}
		\end{figure}
		
		\end{widetext}

		\begin{table*}
		\caption{Marginalized constrained data of the parameters $\omega$, $\alpha$ and $n$ and corresponding anisotropy measure $\Delta$ for different data samples with 68\% confidence level. }
		    \label{tab:my_label}
		\begin{ruledtabular}
		    \centering
		    \begin{tabular}{c c c c c}
				Dataset & $\omega$ & $\alpha$ & n & $\Delta$\\ 
				\hline
				$H(z)$ & $-1.245^{+0.034}_{-0.031}$ & $5.57^{+0.61}_{-0.85}$ & $26.73\pm 0.36$ & $1.604\pm 0.005$\\
				Pantheon & $-1.175^{+0.030}_{-0.020}$  & $6.4\pm 0.93$ & $31.09\pm0.54$ & $1.626^{+0.003}_{-0.004}$\\
				Pantheon+$H(z)$ & $-1.281^{+0.052}_{-0.037}$ & $4.76^{+0.68}_{-0.84}$ & $28.65\pm 0.29$ & $1.654^{+0.005}_{-0.006}$\\
			\end{tabular}
		   \end{ruledtabular}
		\end{table*}

		\subsection{Results}
		\par Heretofore, we have checked over different data samples and have obtained the constraint values for the unknown parameters $\omega$, $\alpha$, and $n$. Further, we obtained the two-dimensional likelihood contours with $1-\sigma$ and $2-\sigma$ errors that are equipped with 68\% and 95\% confidence levels for Hubble, Pantheon, and Hubble+Pantheon data samples. These are depicted in \figurename \ref{fig:kh}, \ref{fig:kp} and \ref{fig:Kcombine}, respectively. First, we considered the $H(z)$ dataset with 57 data points. Here, for the model parameter $\alpha$, we have obtained the value $5.57^{+0.61}_{-0.85}$ and for the parameter $n$, which gives the relation between the directional Hubble parameters, the constrain value turns out to be $26.73\pm 0.36$. Next, for the SNe Ia Pantheon data sets with 1048 sample points, it yields, $\alpha=6.4\pm 0.93$ and $n=31.09\pm0.54$. Finally, for combined data sets in the last section, they attain the values, $\alpha=4.76^{+0.68}_{-0.84}$ and $n=28.65\pm 0.29$. Along with these, to compare our model with the $\Lambda$CDM model, we checked Hubble parameter, $H(z)$ profile and distance modulus, $\mu(z)$ profile  with the constraint values of unknown parameters $\omega$, $\alpha$, and $n$ for $H(z)$ and pantheon samples and illustrated subsequently in figures \ref{fig:khf} and \ref{fig:kpf}. It is observed that, for both cases, our $\mathpzc{f}(\mathcal{R},\mathscr{L}_m)$ model fits nicely with the observational results. Moreover, it is also seen that our model is quite close to the $\Lambda$CDM model's profile. Moreover, it is well-known that the equation of state parameter $\omega$ also plays a crucial role in describing the different energy dominated evolution process of the universe. The present scenario of the universe can predict by either quintessence phase $\left(-1<\omega<-\frac{1}{3}\right)$ or phantom phase $(\omega<-1)$. Now, for the present model, we found $\omega=-1.245^{+0.034}_{-0.031}$, $\omega=-1.175^{+0.030}_{-0.020}$ and $\omega=-1.281^{+0.052}_{-0.037}$, for $H(z)$, Pantheon, and $H(z)$+ Pantheon samples, respectively. Our results on $\omega$ align with the outputs of some observational studies [please see \cite{pantheon,w1,w2}]. It worthy to mention here that, our $\mathpzc{f}(\mathcal{R},\mathscr{L}_m)$ model admits phantom behavior for each data analysis. The contraint values so obtained are summarized in table \ref{tab:my_label}.
		
		\section{CONCLUDING REMARKS}\label{sectionV}
		\par The never-ending curiosity of the scientific community about the current cosmological aspects fosters to look into the universe beyond the standard gravity models. In this direction, $\mathpzc{f}(\mathcal{R},\mathscr{L}_m)$ formalism works pretty well. In the present article, we investigated the accelerated expansion of the universe in the realm of $\mathpzc{f}(\mathcal{R},\mathscr{L}_m)$ gravity. In particular, we adopted a non-linear $\mathpzc{f}(\mathcal{R},\mathscr{L}_m)$ model $\mathpzc{f}(\mathcal{R},\mathscr{L}_m)=\dfrac{1}{2}\mathcal{R}+\mathscr{L}_m^{\,{\alpha}}$. Further, in this, we focused on Bianchi I cosmology which is locally rotationally symmetric. Also, we considered expansion and shear scalar to vary proportionally which can lead to the isotropization of hubble expansion. 
        \par Then, to find the constraint values for the parameters we used the statistical MCMC approach with the Bayesian technique. Further, we analyzed the result for two different observational samples such as Hubble data and Pantheon data( which includes SDSS, SNLS, Pan-STARRS1, low-redshift survey, and HST surveys).  Furthermore, the equation of state parameter, which is significant in explaining the behavior of the universe, has been constrained. The constraint value so-obtained for $\omega$ $\left(-1.245^{+0.034}_{-0.031}, -1.175^{+0.030}_{-0.020}, -1.281^{+0.052}_{-0.037}\right)$, suggests the phantom behavior of the universe. In addition, with these values for the parameters, we compared our model with the $\Lambda$CDM model.
        
        \par Together with this, we can correlate the obtained outcomes with the existing results to assess the present aspects of the universe. Besides, in dealing with a modified theoretic approach, to discuss these scenarios, we use cosmographic treatments and observational constraints. The usage of the former technique has led to numerous interesting investigations within the framework of several modified theories. For instance, appraising the cosmographic parameters such as the deceleration parameter, and equation of state parameter with their present data helps us to examine the cosmic evolution \cite{c1,c2,c3}. Also, the latter approach of observational studies has been extensively done over the past few years \cite{c4,c5,c6,c7,c8}. Moreover, to focus on the equation of state parameter $\omega$, we can see numerous works with a fixed value of $\omega$, say 1/3, 0, -1/3 so on, depending on the fluid dominated in the spacetime. Interestingly, in our work, such supposition for the $\omega$ value has not been admitted. Instead, more advantageously, its value has been constrained against observational results. As per the obtained outcome for $\omega$, we can infer the cosmic acceleration.
        
        \par Finally, for our model, we examined the nature of anisotropy with the aid of the anisotropy parameter. The anisotropy measure for $H(z)$, pantheon, and $H(z)$+pantheon is found as $1.604\pm 0.005$, $1.626^{+0.003}_{-0.004}$ and $1.654^{+0.005}_{-0.006}$ respectively. This model, under all the assumptions made, predict an anisotropy that is in agreement with the dataset used.
        \par In all, these results could motivate us to further explore the studies in $\mathpzc{f}(\mathcal{R},\mathscr{L}_m)$ theory as this obeys the observational data. Moreover, it would be interesting to investigate the inflationary scenario of the universe in the back ground of this theory of gravity. In future, we aim to study this scenario.
        
\section*{Data Availability Statement}
There are no new data associated with this article.

\begin{acknowledgments}
N.S.K. and V.V. acknowledge DST, New Delhi, India, for its financial support for research facilities under DST-FIST-2019. S.M. acknowledges Department of Science \& Technology (DST), Govt. of India, New Delhi, for awarding INSPIRE Fellowship (File No. DST/INSPIRE Fellowship/2018/IF180676). We are very much grateful to the honorable referee and to the editor for the illuminating suggestions that have significantly improved our work in terms of research quality, and presentation.
\end{acknowledgments}

%\nocite{*}

\end{document}